\documentclass[reprint,twocolumn,superscriptaddress,amsmath,amssymb,showpacs,prl]{revtex4-1}
\usepackage{hyperref}
\pdfoutput=1 

\usepackage{dcolumn}
\usepackage{epsfig}
\usepackage{graphics}
\usepackage[latin1]{inputenc} 
\usepackage[T1]{fontenc}
\usepackage{bm}
\usepackage{hyperref}
\newcommand{\ddst}{false}

\begin{document}

\title{Ion-Exchange Strengthening of Glasses: Atomic Topology Matters}

 \author{Mengyi Wang}
 \affiliation{Physics of AmoRphous and Inorganic Solids Laboratory (PARISlab), Department of Civil and Environmental Engineering, University of California, Los Angeles, CA, USA}
\author{Mathieu Bauchy}
 \affiliation{Physics of AmoRphous and Inorganic Solids Laboratory (PARISlab), Department of Civil and Environmental Engineering, University of California, Los Angeles, CA, USA}
 
\date{\today}

\pacs{61.43.-j, 62.40.+i, 62.25.Mn}

\begin{abstract}
Ion-exchange is commonly used to chemically strengthen glasses, by replacing small atoms by larger ones at sub T$_g$ temperature, thereby inducing a compressive stress. However, the resulting expansion of the glass remains lower than that predicted by the difference of molar volumes of the as-cooled glasses, an anomaly that remains poorly understood. Here, based on molecular dynamcis simulations of permanently densified sodium silicate glasses coupled with topological constraint theory, we show that the rigidity of the network controls the extent of the dilatation. Isostatic networks, which are rigid but free of eigenstress, show maximal expansion and, therefore, appear to be an attractive option to improve the toughness of glass.
\end{abstract}

\maketitle

\section{Problem definition: the need for ultra-strong glasses}

Due to their non-crystallinity, glasses show unique properties in terms of optical, electronic and mechanical properties \cite{varshneya_fundamentals_1993}. In particular, the transparency of usual silicate glasses is at the root of life changing applications like optical fibers or smartphones. However, even after decades of intensive research, glasses still break rather easily (smartphone drops often remain fatal) \cite{wang_intrinsic_2015, wang_nano-ductility_2015, yu_fracture_2015}. Their brittleness seriously limit the range of applications, which has been identified as a grand challenge for glasses \cite{mauro_two_2014, mauro_grand_2014}.

Over the last decade, driven by the need for stronger screens for smartphones and tablets, the chemical strengthening of glasses through ion-exchange \cite{krohn_strengthening_1969, cooper_strengthening_1969} has emerged as the easiest and cheapest solution to this problem. The most famous example is probably Corning$^\circledR$ Gorilla$^\circledR$ Glass \cite{wray_gorilla_2013, welch_dynamics_2013}. This process involves the replacement of smaller ions (typically Na$^+$) by larger ones (typically K$^+$) at the surface of the glass, which is usually achieved by placing the glass in a salt bath (typically KNO$_3$) at a temperature below glass transition \cite{wondraczek_towards_2011}. This induces a compressive stress over tens of micrometers of the surface, which limits the risk of cracks propagating from surface flaws when subjected to tension. If potential stress relaxation is neglected, the compressive stress profile obtained by ion-exchange follows:

\begin{equation}
\label{eq:stress}
\sigma = - \frac{B E}{1 - \nu} \left( C(z) - C_{\rm avg}\right)
\end{equation} when $B$ is given by:

\begin{equation}
\label{eq:cooper}
B = \frac{1}{3} \frac{1}{V} \frac{\partial V}{\partial C} 
\end{equation}

where $E$ is the Young's modulus, $\nu$ is the Poisson's ratio, $C(z)$ is the local concentration of substituted cation at a depth $z$ from the surface, $C_{\rm avg}$ is the average of such concentration over the whole range $z$-axis, and $B$ is the linear network dilation coefficient, also known as the Cooper coefficient, in honor of Prof. Alfred R. Cooper \cite{tandia_atomistic_2012}. The linear network dilation coefficient was defined in analogy to the thermal expansion coefficient with respect to the relative change of the molar volume $V$ of the glass.
Despite a large amount of empirical knowledge, the fact that $B$ is typically lower than the value $B_{\rm max}$ that would be expected from the density difference between the as-cooled alkali silicate glasses (e.g., between sodium and potassium silicate) remains poorly understood \cite{varshneya_chemical_2010}. This is a serious limitation, as higher values of $B$ would permit higher surface compressive stresses. This project specifically aims to better understand the role of the rigidity of the atomic network of silicate glasses on the relative amount of strengthening due to ion-exchange.

\begin{figure}
\begin{center}
\includegraphics*[width=\linewidth, keepaspectratio=true, draft=\ddst]{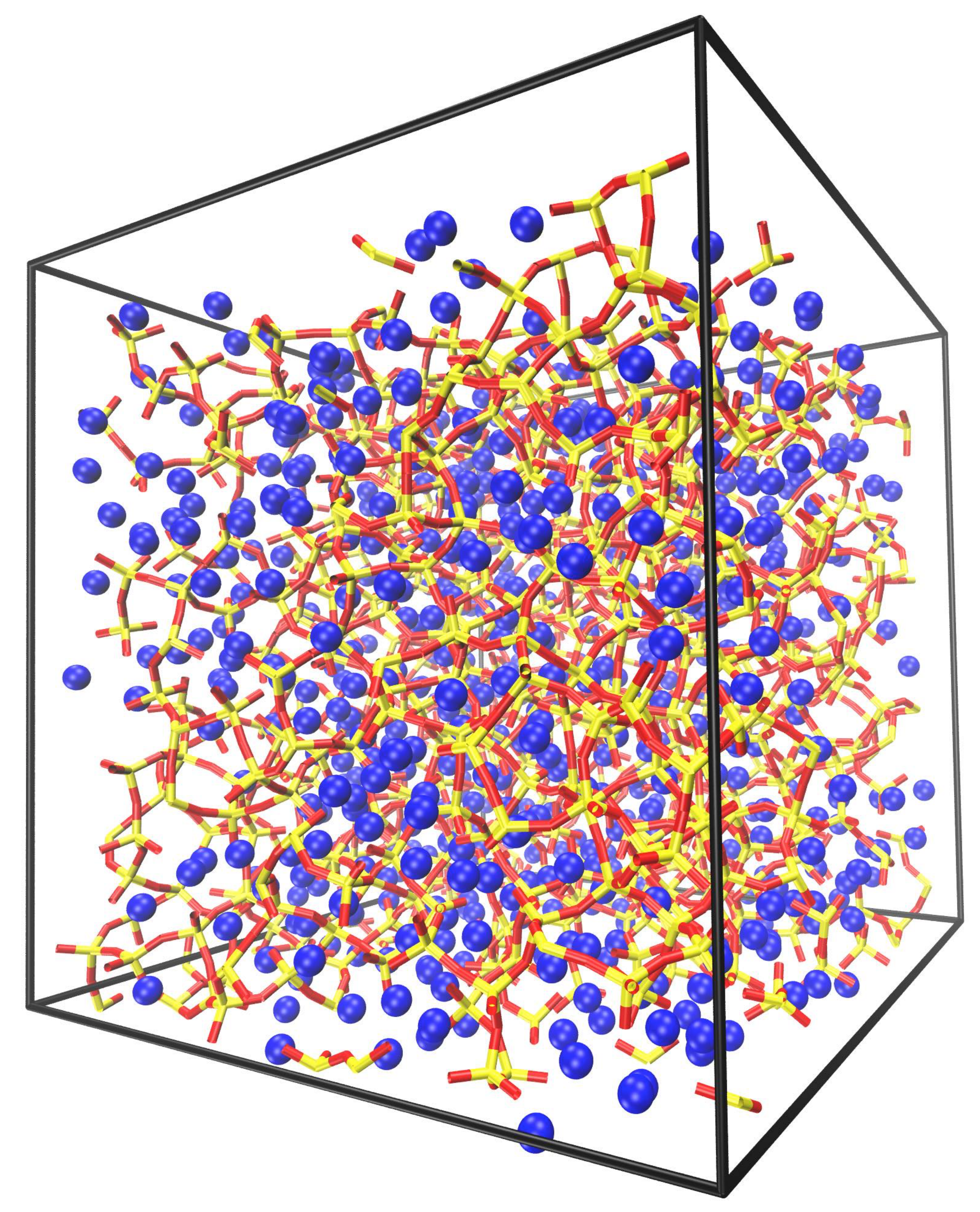}
\caption{\label{fig:snap} Snapshot of the atomic structure of a sodium silicate glass cooled under zero pressure. Si, O, and Na atoms are represented in yellow, red, and blue, respectively.
}
\end{center}	
\end{figure}

\section{Method: molecular dynamics coupled with topological constraints analysis}

As accessing an atomistic picture of the effect of ion-exchange by conventional experimental techniques would be challenging, we rely on classical molecular dynamics (MD) simulations. Such simulations are analyzed within the framework of topological constraints theory (TCT) to characterize the rigidity of the network. TCT considers the complex atomic networks of glasses analogous to simple mechanical trusses while filtering out the chemical details that ultimately do not affect macroscopic properties \cite{mauro_topological_2011, bauchy_topological_2012}. The atoms undergo some mechanical constraints arising from chemical interactions, which can be divided into two types: the 2-body radial bond-stretching constraints and the 3-body angular bond-bending constraints. Following Maxwell's criterion of stability for mechanical trusses \cite{maxwell_l._1864}, glassy networks can either be flexible, stressed-rigid, or isostatic, when the number of constraints per atom, $n_{\rm c}$, is lower, higher, or equal, respectively, to 3, the number of degrees of freedom per atom.

\begin{figure}
\begin{center}
\includegraphics*[width=\linewidth, keepaspectratio=true, draft=\ddst]{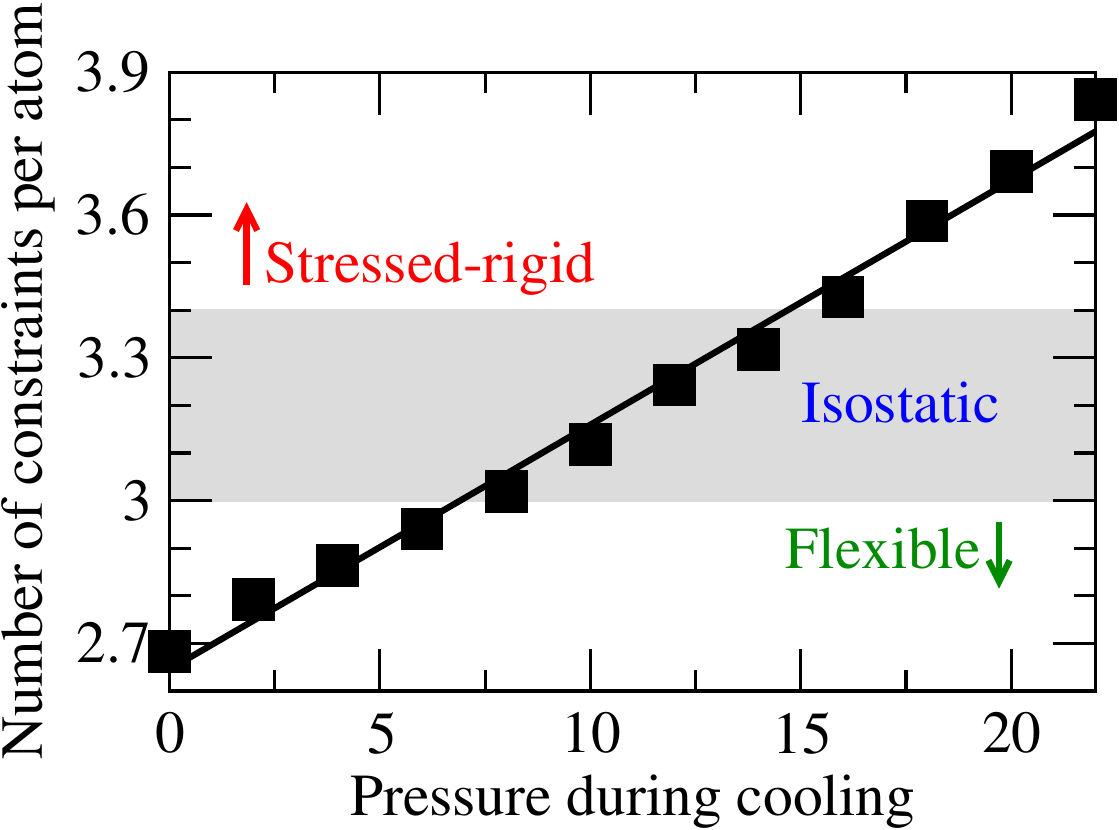}
\caption{\label{fig:nc} Number of constraints per atom $n_{\rm c}$, for permanently densified sodium silicate glasses, with respect to the pressure applied during the cooling phase. The grey area indicates the domain of maximum expansion due to ion-exchange.
}
\end{center}	
\end{figure}

To elucidate the influence of atomic rigidity on ion-exchange, sodium silicate glasses (Na$_2$O)$_{30}$(SiO$_2$)$_{70}$ (noted NS hereafter, 3000 atoms, see Fig. \ref{fig:snap}) cooled under constant pressure $P$ are prepared by MD, followed by relaxation to zero pressure at room temperature, but remaining permanently densified \cite{bauchy_viscosity_2013, bauchy_structural_2012, bauchy_pockets_2011}. As shown in Fig. \ref{fig:nc}, these glasses have been reported to show a rigidity transition with respect to $P$: flexible at low $P$ and stressed-rigid at high $P$ due to the change of the coordination number of Si and O atoms. This effectively delimits an isostatic pressure window around 8 GPa, inside which NS shows a reversible glass transition \cite{bauchy_densified_2015, bauchy_transport_2013, micoulaut_topological_2015, bauchy_percolative_2013, micoulaut_anomalies_2013}. Typically, glasses show rigidity transitions according to composition, but a change of composition (e.g., the fraction of Na, or a replacement of Si by Al atoms) would affect alkali atoms and therefore ion-exchange strengthening, which would not allow us to identify the effect of network topology \textit{only}.

\section{Results: computed ion-exchange compressive stress}

Taking the permanently densified NS systems as host glasses, we simulate the effect of ion-exchange by manually replacing a given fraction of Na$^+$ by K$^+$, following a methodology already well-established \cite{tandia_atomistic_2012, vargheese_molecular_2014}. After ion-exchange, the glass is relaxed at zero pressure in order to compute the resulting increase of volume. All the processes are performed at 300K in order to avoid stress relaxation \cite{sane_stress_1987}.

\begin{figure}
\begin{center}
\includegraphics*[width=\linewidth, keepaspectratio=true, draft=\ddst]{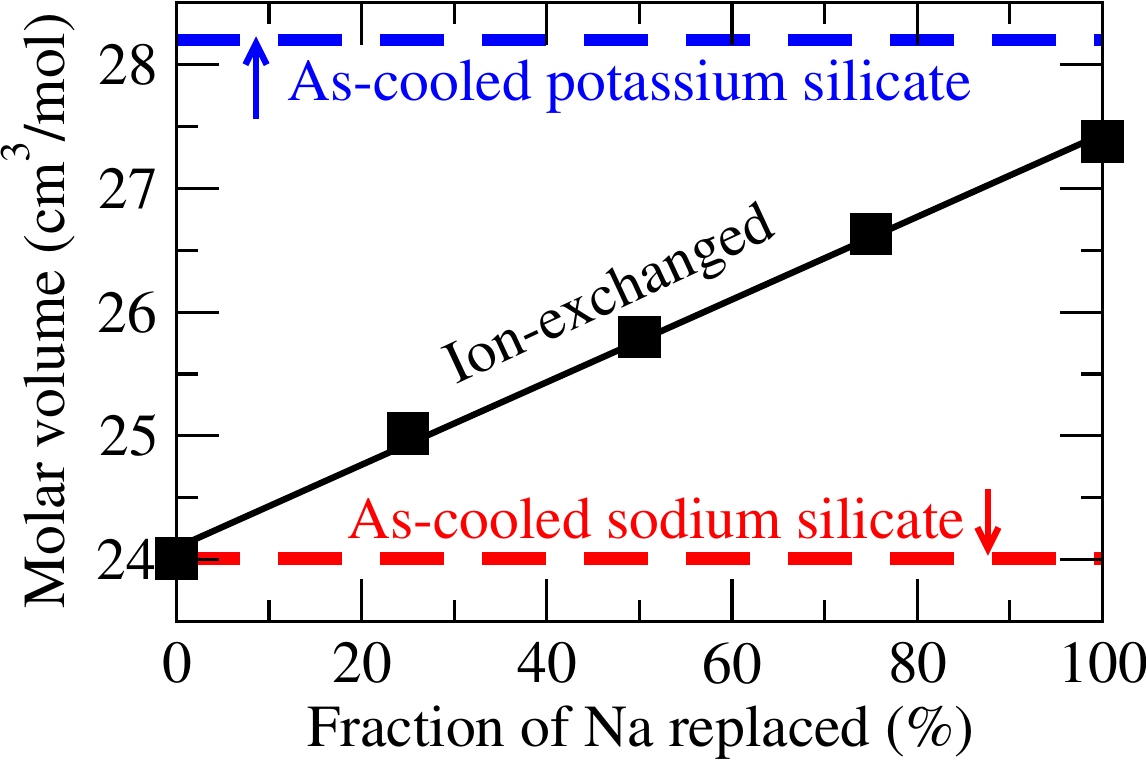}
\caption{\label{fig:Vm} Molar volume of a sodium silicate glass, cooled under zero pressure, after replacement of a given fraction of Na$^+$ by K$^+$ ions. The red (blue) dashed line indicates the molar volume of as-cooled sodium (potassium) silicate.
}
\end{center}	
\end{figure}

Fig. \ref{fig:Vm} shows the computed molar volume of ion-exchanged sodium silicate (cooled under zero pressure) with respect to the fraction of Na$^+$ replaced by K$^+$ ions. As suggested by Eq. \ref{eq:stress}, the molar volume of the ion-exchanged glasses increases fairly linearly with the fraction of substituted cations. We note that, similar to what is observed in alkali aluminosilicate glasses \cite{vargheese_molecular_2014}, once all Na$^+$ have been replaced by K$^+$ ions, the molar volume of the ion-exchanged glass remains lower than that of the as-cooled potassium silicate glass (noted KS hereafter). This has been attributed to the fact that the ion-exchanged system is a \textit{forbidden glass} \cite{mauro_forbidden_2009} in which K atoms have a different coordination number than that in KS glass \cite{tandia_atomistic_2012}. However, we observe that the molar volume of ion-exchanged NS glasses reaches 97\% of that of as-cooled KS glass, whereas, using the same methodology, the molar volume of ion-exchanged (Na$_2$O)$_{20}$(Al$_2$O$_3$)$_{20}$(SiO$_2$)$_{60}$ glass reaches only 94\% of that of the corresponding as-cooled potassium aluminosilicate glass \cite{vargheese_molecular_2014}. This highlights the fact that network formers atoms and, more generally, the rigidity of the network influence the extent of volume expansion induced by ion-exchange.

\begin{figure}
\begin{center}
\includegraphics*[width=\linewidth, keepaspectratio=true, draft=\ddst]{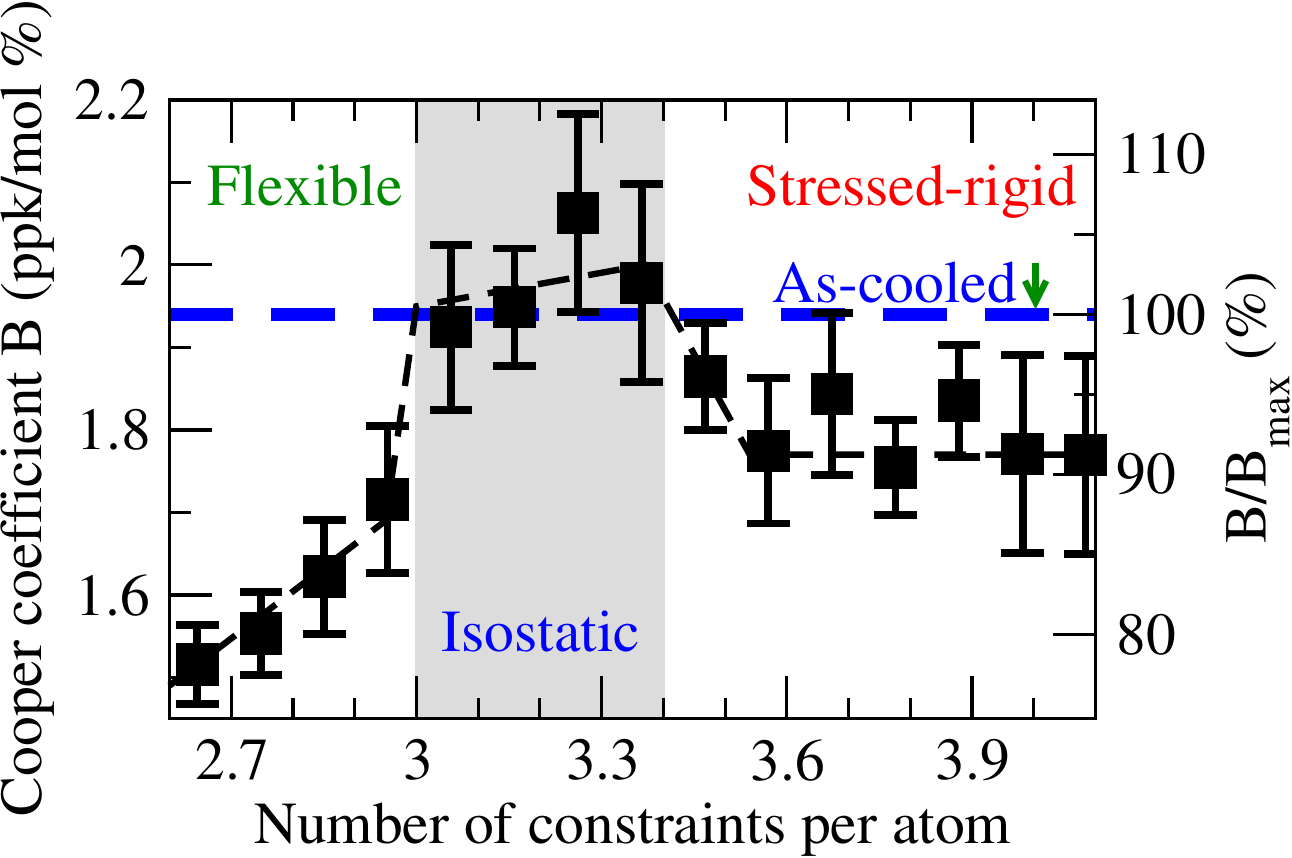}
\caption{\label{fig:B} Cooper (linear network dilation) coefficient $B$ after substitution of all Na$^+$ by K$^+$ ions, with respect to the number of constraints per atom $n_{\rm c}$. The grey area indicates the domain of maximum expansion due to ion-exchange. The blue dashed line indicates $B_{\rm max}$, as obtained from the molar volume of as-cooled sodium and potassium silicate. The black dashed line serves as a guide for the eye.
}
\end{center}	
\end{figure}

Following this idea, we perform the same analysis on permanently densified NS glass with different degrees of rigidity (see Fig. \ref{fig:nc}) by replacing 100\% of Na$^+$ by K$^+$ in each case, and monitoring the resulting increase of molar volume. Fig. \ref{fig:B} shows the Cooper coefficient $B$, as calculated from Eq. \ref{eq:cooper}, with respect to the number of constraints per atom $n_{\rm c}$ before ion-exchange in the host NS glasses. Interestingly, $B$ shows a broad maximum within $n_{\rm c}$ = 3-3.4, which corresponds to a window of pressure during cooling of 8-16 GPa. This pressure window features a maximum of diffusion in the super-cooled liquid state \cite{bauchy_transport_2013} and has been shown to be characterized by an isostatic network \cite{bauchy_densified_2015}. We note that, in the pressure window, the Cooper coefficient reaches its theoretical maximum value, as obtained by the difference of molar volume between the as-cooled NS and KS glasses, which remains fairly constant with different cooling pressures.

\section{Discussion: competition between rigidity and stress}

\begin{figure*}
\begin{center}
\includegraphics*[width=\linewidth, keepaspectratio=true, draft=\ddst]{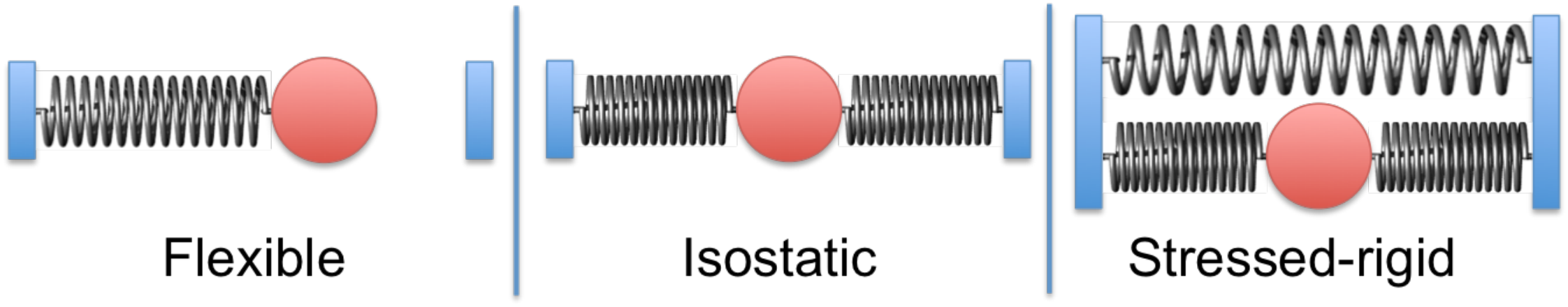}
\caption{\label{fig:springs} Schematic of the origin of the maximum compressive stress in isostatic networks.
}
\end{center}	
\end{figure*}

Question remains about the origin of this maximum in the isostatic phase. We propose the following atomistic picture (as illustrated in Fig. \ref{fig:springs}). (1) Flexible networks feature internal modes of deformation, or floppy modes \cite{jacobs_generic_1995}, which allow for local deformations with low energy barriers. The stress imposed by the stuffing of larger atoms can therefore be partially absorbed by local reorganizations of the network. (2) On the contrary, stressed-rigid networks are completely locked and cannot easily deform. Hence, rather than showing a global expansion, the induced stress is distributed as eigenstress \cite{wang_pressure_2005} within the network where some of the bonds are stretched and some others are compressed, as shown in Fig. \ref{fig:springs}. (3) Eventually, isostatic networks are characterized by a percolation of rigidity \cite{jacobs_generic_1995} so that the displacement of one atom should affect the entire network, but are free of eigenstress. Consequently, they feature the largest expansion upon ion-exchange. All together, this suggests that a better tuning of the rigidity of the host glass would allow us to reach higher compressive stress and, therefore, to design stronger glasses.

\section{Future work}

For further research, we will try to find the structural signature of isostatic glasses to understand why they strengthen in an optimal way. This structural signature could be used as a metrics to quickly identify optimal compositions, e.g., from diffraction patterns. Although pure sodium silicate glass offers a simple starting point for us to better understand the underlying physics of ion-exchange, such a study will be extended to more realistic compositions of glass, e.g., modified aluminosilicate glasses. However, the rigidity of such glasses, so far unknown, will have to be evaluated first. Also, we keep in mind that higher surface compressive stress could also be compensated by higher stress relaxation. Although such relaxations are challenging to study directly via MD, they can potentially be predicted by accelerated relaxation techniques \cite{yu_stretched_2015}. This option will be evaluated in the future.

\begin{acknowledgments}
The authors acknowledge financial support for this research provisioned by the University of California, Los Angeles (UCLA). Access to computational resources was provisioned by the Physics of AmoRphous and Inorganic Solids Laboratory (PARISlab).
\end{acknowledgments}


\begin{thebibliography}{30}
\expandafter\ifx\csname natexlab\endcsname\relax\def\natexlab#1{#1}\fi
\expandafter\ifx\csname bibnamefont\endcsname\relax
  \def\bibnamefont#1{#1}\fi
\expandafter\ifx\csname bibfnamefont\endcsname\relax
  \def\bibfnamefont#1{#1}\fi
\expandafter\ifx\csname citenamefont\endcsname\relax
  \def\citenamefont#1{#1}\fi
\expandafter\ifx\csname url\endcsname\relax
  \def\url#1{\texttt{#1}}\fi
\expandafter\ifx\csname urlprefix\endcsname\relax\def\urlprefix{URL }\fi
\providecommand{\bibinfo}[2]{#2}
\providecommand{\eprint}[2][]{\url{#2}}

\bibitem[{\citenamefont{Varshneya}(1993)}]{varshneya_fundamentals_1993}
\bibinfo{author}{\bibfnamefont{A.~K.} \bibnamefont{Varshneya}},
  \emph{\bibinfo{title}{Fundamentals of {Inorganic} {Glasses}}}

\bibitem[{\citenamefont{Wang et~al.}(2015{\natexlab{a}})\citenamefont{Wang, Yu,
  Lee, and Bauchy}}]{wang_intrinsic_2015}
\bibinfo{author}{\bibfnamefont{B.}~\bibnamefont{Wang}},
  \bibinfo{author}{\bibfnamefont{Y.}~\bibnamefont{Yu}},
  \bibinfo{author}{\bibfnamefont{Y.~J.} \bibnamefont{Lee}}, \bibnamefont{and}
  \bibinfo{author}{\bibfnamefont{M.}~\bibnamefont{Bauchy}},
  \bibinfo{journal}{Glass Science} \textbf{\bibinfo{volume}{2}},
  \bibinfo{pages}{11} (\bibinfo{year}{2015}{\natexlab{a}}).

\bibitem[{\citenamefont{Wang et~al.}(2015{\natexlab{b}})\citenamefont{Wang, Yu,
  Wang, and Bauchy}}]{wang_nano-ductility_2015}
\bibinfo{author}{\bibfnamefont{B.}~\bibnamefont{Wang}},
  \bibinfo{author}{\bibfnamefont{Y.}~\bibnamefont{Yu}},
  \bibinfo{author}{\bibfnamefont{M.}~\bibnamefont{Wang}}, \bibnamefont{and}
  \bibinfo{author}{\bibfnamefont{M.}~\bibnamefont{Bauchy}},
  \bibinfo{journal}{arXiv:1505.04486 [cond-mat]}
  (\bibinfo{year}{2015}{\natexlab{b}}).

\bibitem[{\citenamefont{Yu et~al.}(2015{\natexlab{a}})\citenamefont{Yu, Wang,
  Lee, and Bauchy}}]{yu_fracture_2015}
\bibinfo{author}{\bibfnamefont{Y.}~\bibnamefont{Yu}},
  \bibinfo{author}{\bibfnamefont{B.}~\bibnamefont{Wang}},
  \bibinfo{author}{\bibfnamefont{Y.~J.} \bibnamefont{Lee}}, \bibnamefont{and}
  \bibinfo{author}{\bibfnamefont{M.}~\bibnamefont{Bauchy}}, in
  \emph{\bibinfo{booktitle}{Symposium {UU} {\textendash} {Structure}-{Property}
  {Relations} in {Amorphous} {Solids}}} (\bibinfo{year}{2015}{\natexlab{a}}),

\bibitem[{\citenamefont{Mauro and Zanotto}(2014)}]{mauro_two_2014}
\bibinfo{author}{\bibfnamefont{J.~C.} \bibnamefont{Mauro}} \bibnamefont{and}
  \bibinfo{author}{\bibfnamefont{E.~D.} \bibnamefont{Zanotto}},
  \bibinfo{journal}{International Journal of Applied Glass Science}
  \textbf{\bibinfo{volume}{5}}, \bibinfo{pages}{313} (\bibinfo{year}{2014}).

\bibitem[{\citenamefont{Mauro}(2014)}]{mauro_grand_2014}
\bibinfo{author}{\bibfnamefont{J.~C.} \bibnamefont{Mauro}},
  \bibinfo{journal}{Glass Science} \textbf{\bibinfo{volume}{1}},
  \bibinfo{pages}{20} (\bibinfo{year}{2014}).

\bibitem[{\citenamefont{Krohn and Cooper}(1969)}]{krohn_strengthening_1969}
\bibinfo{author}{\bibfnamefont{D.~A.} \bibnamefont{Krohn}} \bibnamefont{and}
  \bibinfo{author}{\bibfnamefont{A.~R.} \bibnamefont{Cooper}},
  \bibinfo{journal}{Journal of the American Ceramic Society}
  \textbf{\bibinfo{volume}{52}}, \bibinfo{pages}{661} (\bibinfo{year}{1969}).

\bibitem[{\citenamefont{Cooper and Krohn}(1969)}]{cooper_strengthening_1969}
\bibinfo{author}{\bibfnamefont{A.~R.} \bibnamefont{Cooper}} \bibnamefont{and}
  \bibinfo{author}{\bibfnamefont{D.~A.} \bibnamefont{Krohn}},
  \bibinfo{journal}{Journal of the American Ceramic Society}
  \textbf{\bibinfo{volume}{52}}, \bibinfo{pages}{665} (\bibinfo{year}{1969}).

\bibitem[{\citenamefont{Wray}(2013)}]{wray_gorilla_2013}
\bibinfo{author}{\bibfnamefont{P.}~\bibnamefont{Wray}},
  \bibinfo{journal}{Ceramic Tech Today}  (\bibinfo{year}{2013}).

\bibitem[{\citenamefont{Welch et~al.}(2013)\citenamefont{Welch, Smith, Potuzak,
  Guo, Bowden, Kiczenski, Allan, King, Ellison, and
  Mauro}}]{welch_dynamics_2013}
\bibinfo{author}{\bibfnamefont{R.}~\bibnamefont{Welch}},
  \bibinfo{author}{\bibfnamefont{J.}~\bibnamefont{Smith}},
  \bibinfo{author}{\bibfnamefont{M.}~\bibnamefont{Potuzak}},
  \bibinfo{author}{\bibfnamefont{X.}~\bibnamefont{Guo}},
  \bibinfo{author}{\bibfnamefont{B.}~\bibnamefont{Bowden}},
  \bibinfo{author}{\bibfnamefont{T.}~\bibnamefont{Kiczenski}},
  \bibinfo{author}{\bibfnamefont{D.}~\bibnamefont{Allan}},
  \bibinfo{author}{\bibfnamefont{E.}~\bibnamefont{King}},
  \bibinfo{author}{\bibfnamefont{A.}~\bibnamefont{Ellison}}, \bibnamefont{and}
  \bibinfo{author}{\bibfnamefont{J.}~\bibnamefont{Mauro}},
  \bibinfo{journal}{Physical Review Letters} \textbf{\bibinfo{volume}{110}},
  \bibinfo{pages}{265901} (\bibinfo{year}{2013}).

\bibitem[{\citenamefont{Wondraczek et~al.}(2011)\citenamefont{Wondraczek,
  Mauro, Eckert, K{\"u}hn, Horbach, Deubener, and
  Rouxel}}]{wondraczek_towards_2011}
\bibinfo{author}{\bibfnamefont{L.}~\bibnamefont{Wondraczek}},
  \bibinfo{author}{\bibfnamefont{J.~C.} \bibnamefont{Mauro}},
  \bibinfo{author}{\bibfnamefont{J.}~\bibnamefont{Eckert}},
  \bibinfo{author}{\bibfnamefont{U.}~\bibnamefont{K{\"u}hn}},
  \bibinfo{author}{\bibfnamefont{J.}~\bibnamefont{Horbach}},
  \bibinfo{author}{\bibfnamefont{J.}~\bibnamefont{Deubener}}, \bibnamefont{and}
  \bibinfo{author}{\bibfnamefont{T.}~\bibnamefont{Rouxel}},
  \bibinfo{journal}{Advanced Materials} \textbf{\bibinfo{volume}{23}},
  \bibinfo{pages}{4578} (\bibinfo{year}{2011}).

\bibitem[{\citenamefont{Tandia et~al.}(2012)\citenamefont{Tandia, Vargheese,
  Mauro, and Varshneya}}]{tandia_atomistic_2012}
\bibinfo{author}{\bibfnamefont{A.}~\bibnamefont{Tandia}},
  \bibinfo{author}{\bibfnamefont{K.~D.} \bibnamefont{Vargheese}},
  \bibinfo{author}{\bibfnamefont{J.~C.} \bibnamefont{Mauro}}, \bibnamefont{and}
  \bibinfo{author}{\bibfnamefont{A.~K.} \bibnamefont{Varshneya}},
  \bibinfo{journal}{Journal of Non-Crystalline Solids}
  \textbf{\bibinfo{volume}{358}}, \bibinfo{pages}{316} (\bibinfo{year}{2012}).

\bibitem[{\citenamefont{Varshneya}(2010)}]{varshneya_chemical_2010}
\bibinfo{author}{\bibfnamefont{A.~K.} \bibnamefont{Varshneya}},
  \bibinfo{journal}{International Journal of Applied Glass Science}
  \textbf{\bibinfo{volume}{1}}, \bibinfo{pages}{131} (\bibinfo{year}{2010}).

\bibitem[{\citenamefont{Mauro}(2011)}]{mauro_topological_2011}
\bibinfo{author}{\bibfnamefont{J.~C.} \bibnamefont{Mauro}},
  \bibinfo{journal}{American Ceramic Society Bulletin}
  \textbf{\bibinfo{volume}{90}}, \bibinfo{pages}{31} (\bibinfo{year}{2011}).

\bibitem[{\citenamefont{Bauchy}(2012{\natexlab{a}})}]{bauchy_topological_2012}
\bibinfo{author}{\bibfnamefont{M.}~\bibnamefont{Bauchy}},
  \bibinfo{journal}{American Ceramic Society Bulletin}
  \textbf{\bibinfo{volume}{91}}, \bibinfo{pages}{34}
  (\bibinfo{year}{2012}{\natexlab{a}}).

\bibitem[{\citenamefont{Maxwell}(1864)}]{maxwell_l._1864}
\bibinfo{author}{\bibfnamefont{J.~C.} \bibnamefont{Maxwell}},
  \bibinfo{journal}{Philosophical Magazine Series 4}
  \textbf{\bibinfo{volume}{27}}, \bibinfo{pages}{294} (\bibinfo{year}{1864}).

\bibitem[{\citenamefont{Bauchy et~al.}(2013)\citenamefont{Bauchy, Guillot,
  Micoulaut, and Sator}}]{bauchy_viscosity_2013}
\bibinfo{author}{\bibfnamefont{M.}~\bibnamefont{Bauchy}},
  \bibinfo{author}{\bibfnamefont{B.}~\bibnamefont{Guillot}},
  \bibinfo{author}{\bibfnamefont{M.}~\bibnamefont{Micoulaut}},
  \bibnamefont{and} \bibinfo{author}{\bibfnamefont{N.}~\bibnamefont{Sator}},
  \bibinfo{journal}{Chemical Geology} \textbf{\bibinfo{volume}{346}},
  \bibinfo{pages}{47} (\bibinfo{year}{2013}).

\bibitem[{\citenamefont{Bauchy}(2012{\natexlab{b}})}]{bauchy_structural_2012}
\bibinfo{author}{\bibfnamefont{M.}~\bibnamefont{Bauchy}}, \bibinfo{journal}{The
  Journal of Chemical Physics} \textbf{\bibinfo{volume}{137}},
  \bibinfo{pages}{044510} (\bibinfo{year}{2012}{\natexlab{b}}).

\bibitem[{\citenamefont{Bauchy and Micoulaut}(2011)}]{bauchy_pockets_2011}
\bibinfo{author}{\bibfnamefont{M.}~\bibnamefont{Bauchy}} \bibnamefont{and}
  \bibinfo{author}{\bibfnamefont{M.}~\bibnamefont{Micoulaut}},
  \bibinfo{journal}{Physical Review B} \textbf{\bibinfo{volume}{83}},
  \bibinfo{pages}{184118} (\bibinfo{year}{2011}).

\bibitem[{\citenamefont{Bauchy and Micoulaut}(2015)}]{bauchy_densified_2015}
\bibinfo{author}{\bibfnamefont{M.}~\bibnamefont{Bauchy}} \bibnamefont{and}
  \bibinfo{author}{\bibfnamefont{M.}~\bibnamefont{Micoulaut}},
  \bibinfo{journal}{Nature Communications} \textbf{\bibinfo{volume}{6}}
  (\bibinfo{year}{2015}).

\bibitem[{\citenamefont{Bauchy and
  Micoulaut}(2013{\natexlab{a}})}]{bauchy_transport_2013}
\bibinfo{author}{\bibfnamefont{M.}~\bibnamefont{Bauchy}} \bibnamefont{and}
  \bibinfo{author}{\bibfnamefont{M.}~\bibnamefont{Micoulaut}},
  \bibinfo{journal}{Physical Review Letters} \textbf{\bibinfo{volume}{110}},
  \bibinfo{pages}{095501} (\bibinfo{year}{2013}{\natexlab{a}}).

\bibitem[{\citenamefont{Micoulaut et~al.}(2015)\citenamefont{Micoulaut, Bauchy,
  and Flores-Ruiz}}]{micoulaut_topological_2015}
\bibinfo{author}{\bibfnamefont{M.}~\bibnamefont{Micoulaut}},
  \bibinfo{author}{\bibfnamefont{M.}~\bibnamefont{Bauchy}}, \bibnamefont{and}
  \bibinfo{author}{\bibfnamefont{H.}~\bibnamefont{Flores-Ruiz}}, in
  \emph{\bibinfo{booktitle}{Molecular {Dynamics} {Simulations} of {Disordered}
  {Materials}}}, edited by
  \bibinfo{editor}{\bibfnamefont{C.}~\bibnamefont{Massobrio}},
  \bibinfo{editor}{\bibfnamefont{J.}~\bibnamefont{Du}},
  \bibinfo{editor}{\bibfnamefont{M.}~\bibnamefont{Bernasconi}},
  \bibnamefont{and} \bibinfo{editor}{\bibfnamefont{P.~S.} \bibnamefont{Salmon}}
  (\bibinfo{publisher}{Springer International Publishing},
  \bibinfo{year}{2015}), no. \bibinfo{number}{215} in \bibinfo{series}{Springer
  {Series} in {Materials} {Science}},

\bibitem[{\citenamefont{Bauchy and
  Micoulaut}(2013{\natexlab{b}})}]{bauchy_percolative_2013}
\bibinfo{author}{\bibfnamefont{M.}~\bibnamefont{Bauchy}} \bibnamefont{and}
  \bibinfo{author}{\bibfnamefont{M.}~\bibnamefont{Micoulaut}},
  \bibinfo{journal}{EPL (Europhysics Letters)} \textbf{\bibinfo{volume}{104}},
  \bibinfo{pages}{56002} (\bibinfo{year}{2013}{\natexlab{b}}).

\bibitem[{\citenamefont{Micoulaut and Bauchy}(2013)}]{micoulaut_anomalies_2013}
\bibinfo{author}{\bibfnamefont{M.}~\bibnamefont{Micoulaut}} \bibnamefont{and}
  \bibinfo{author}{\bibfnamefont{M.}~\bibnamefont{Bauchy}},
  \bibinfo{journal}{physica status solidi (b)} \textbf{\bibinfo{volume}{250}},
  \bibinfo{pages}{976} (\bibinfo{year}{2013}).

\bibitem[{\citenamefont{Vargheese et~al.}(2014)\citenamefont{Vargheese, Tandia,
  and Mauro}}]{vargheese_molecular_2014}
\bibinfo{author}{\bibfnamefont{K.~D.} \bibnamefont{Vargheese}},
  \bibinfo{author}{\bibfnamefont{A.}~\bibnamefont{Tandia}}, \bibnamefont{and}
  \bibinfo{author}{\bibfnamefont{J.~C.} \bibnamefont{Mauro}},
  \bibinfo{journal}{Journal of Non-Crystalline Solids}
  \textbf{\bibinfo{volume}{403}}, \bibinfo{pages}{107} (\bibinfo{year}{2014}).

\bibitem[{\citenamefont{Sane and Cooper}(1987)}]{sane_stress_1987}
\bibinfo{author}{\bibfnamefont{A.~Y.} \bibnamefont{Sane}} \bibnamefont{and}
  \bibinfo{author}{\bibfnamefont{A.~R.} \bibnamefont{Cooper}},
  \bibinfo{journal}{Journal of the American Ceramic Society}
  \textbf{\bibinfo{volume}{70}}, \bibinfo{pages}{86} (\bibinfo{year}{1987}).

\bibitem[{\citenamefont{Mauro and Loucks}(2009)}]{mauro_forbidden_2009}
\bibinfo{author}{\bibfnamefont{J.~C.} \bibnamefont{Mauro}} \bibnamefont{and}
  \bibinfo{author}{\bibfnamefont{R.~J.} \bibnamefont{Loucks}},
  \bibinfo{journal}{Journal of Non-Crystalline Solids}
  \textbf{\bibinfo{volume}{355}}, \bibinfo{pages}{676} (\bibinfo{year}{2009}).

\bibitem[{\citenamefont{Jacobs and Thorpe}(1995)}]{jacobs_generic_1995}
\bibinfo{author}{\bibfnamefont{D.~J.} \bibnamefont{Jacobs}} \bibnamefont{and}
  \bibinfo{author}{\bibfnamefont{M.~F.} \bibnamefont{Thorpe}},
  \bibinfo{journal}{Physical Review Letters} \textbf{\bibinfo{volume}{75}},
  \bibinfo{pages}{4051} (\bibinfo{year}{1995}).

\bibitem[{\citenamefont{Wang et~al.}(2005)\citenamefont{Wang, Mamedov,
  Boolchand, Goodman, and Chandrasekhar}}]{wang_pressure_2005}
\bibinfo{author}{\bibfnamefont{F.}~\bibnamefont{Wang}},
  \bibinfo{author}{\bibfnamefont{S.}~\bibnamefont{Mamedov}},
  \bibinfo{author}{\bibfnamefont{P.}~\bibnamefont{Boolchand}},
  \bibinfo{author}{\bibfnamefont{B.}~\bibnamefont{Goodman}}, \bibnamefont{and}
  \bibinfo{author}{\bibfnamefont{M.}~\bibnamefont{Chandrasekhar}},
  \bibinfo{journal}{Physical Review B} \textbf{\bibinfo{volume}{71}}
  (\bibinfo{year}{2005}).

\bibitem[{\citenamefont{Yu et~al.}(2015{\natexlab{b}})\citenamefont{Yu, Wang,
  Zhang, Wang, Sant, and Bauchy}}]{yu_stretched_2015}
\bibinfo{author}{\bibfnamefont{Y.}~\bibnamefont{Yu}},
  \bibinfo{author}{\bibfnamefont{M.}~\bibnamefont{Wang}},
  \bibinfo{author}{\bibfnamefont{D.}~\bibnamefont{Zhang}},
  \bibinfo{author}{\bibfnamefont{B.}~\bibnamefont{Wang}},
  \bibinfo{author}{\bibfnamefont{G.}~\bibnamefont{Sant}}, \bibnamefont{and}
  \bibinfo{author}{\bibfnamefont{M.}~\bibnamefont{Bauchy}},
  \bibinfo{journal}{arXiv:1503.07242 [cond-mat]}
  (\bibinfo{year}{2015}{\natexlab{b}}).

\end{thebibliography}

\end{document}